%381.tex
\input amstex
\documentstyle{amsppt}
%\NoPageNumbers
\NoRunningHeads

\hsize 6.25truein
\hfuzz=25pt
\vsize 9.0truein

\TagsOnRight
Jour. of Inverse and Ill-Posed Problems, 6, N5, (1998), 515-520.
\topmatter
\title On the theory of reproducing kernel Hilbert spaces
\endtitle
%\rightheadtext{ SHORT TITLE }
\author  A.G. Ramm      \endauthor
\affil Department of Mathematics, Kansas State University, 
        Manhattan, KS  66506-2602, USA \\
        ramm\@math.ksu.edu  \\
        \endaffil

%\address \endaddress \email\endemail

\abstract {
The inner product in RKHS is described in abstract form.
Some of the results, published earlier, are discussed from a general
point of view. In particular, the characterization of the
range of linear integral transforms and inversion formulas,
announced in the works of Saitoh, are analyzed.
} \endabstract

\endtopmatter

\vglue .1in

\document

\subhead 1. Introduction \endsubhead

\vskip .1in

The theory of reproducing kernels was developed in [A], [S].
A recent review of the theory is [Sa1,Sa2], where the reader can find
many references.

The basic result in [A] is the existence and uniqueness of a reproducing
kernel Hilbert space (RKHS) corresponding to any
{\it self-adjoint nonnegative-definite kernel}
$K(p,q)$, $p,q\in E$, where $E$ is an abstract set.
Let $H$ be a Hilbert space of functions defined on $E$, and
$H\subset L^2(E)$. Assume that $K(\cdot,q)$ and $K(p,\cdot)$
belong to $H$.
Let us assume that the linear operator $K:H\to H$, with the kernel
$K(p,q)$, is injective.
It is defined on all of $H$ since $K(p,\cdot)\in H$ by the assumption.
Define RKHS $H_K$ inner product by the formula
$$ (f,g)_{H_K}:= [f,g]:=(K^{-1}f,g),
   \tag1.1 $$
where $(f,g):=(f,g)_{L^2(E)}$,  $K^{-1}$ is the operator inverse to
$K:H\to H$, and
$$ Kf:=\int_EK(p,q)f(q)dq.
   \tag 1.2$$
The injectivity assumption can be dropped, but then one has to consider
$K$ on the factor space $H/N(K)$, where $N(K):=\{f:Kf=0\}$
is the null-space of $K$.

In the literature (e.g. see [Sa1,2]) the inner product in RKHS was not
defined explicitly by formula (1.1). The definition of the
inner product in $H_K$, given in [A] (and
presented in [Sa1, p.36]) is
implicit and contains some limiting procedure which is not
described explicitly. In particular, it is not clear over which
sets of $p$ and $q$ the summation in formula (11) in [Sa1, p.36]
is taken. In [A] such a summation is taken over a finite set
of points $p\in E$ and $q\in E$. The finite sums
$\sum_p X_p K(\cdot,p)$, used in [Sa1, p.36] do not form
a complete Hilbert space $H_K$, and the
completion procedure is not discussed
in sufficient details in [Sa1]. Our definition (1.1) of the
inner product in $H_K$ coincides with the definition
in [Sa1, p.36, formula (11)] if one takes $f$ and $g$
in (1.1) to be finite linear combinations of the
functions of the type $K(p,\cdot)$ and $K(\cdot, q)$.

The reproducing property of the kernel $K(p,q)$ can be stated as follows:
$$ [f(\cdot), K(\cdot,q)]=f(q),
   \tag1.3 $$
and this formula can be easily derived from the definition (1.1)
of the inner product in $H_K$:
$$ [f(\cdot),K(\cdot,q)]:=(K^{-1}f,K(\cdot, q))=(f,K^{-1}K(\cdot,q))
   =(f,I(\cdot,q))=f(q).
   \tag1.4 $$
Here we have used the selfadjointness of the operator $K^{-1}$,
and the fact that the distributional kernel of the identity operator
$I$ is $\delta(p-q)$, the delta function, which is well defined
on RKHS because the value $f(p)$ for any $p\in E$ is a bounded
 linear functional in $H$:
$$ |f(q)|\leq ||f||\,||K(\cdot,q)||,
   \tag1.5 $$
where $||f||:=[f,f]^{\frac12}$ is the norm in $H_K$.

The basic results of this paper are: 

1) representation of the inner product in $H_K$ by
formula (1.1),

and

2) clarification of the conditions from [Sa1,2] under which the
range of the general linear transform, defined by
formula (2.1) below, is characterized and inversion formulas
for this transform are obtained.

\subhead 2. Linear transforms and RKHS \endsubhead

Define
$$ f(p):=LF:=\int_T\overline{h(t,p)} F(t) dm(t),
   \tag2.1 $$
where $T\subset {\Bbb R}^n$ is some subset of ${\Bbb R}^n$,
$dm(t)$ is a positive measure on $T$, and $h(t,p)$ is a function on
$H_0\times H$, where $H_0:=L^2(T,dm(t))$.
The linear operator $L:H_0\to H$ is injective if the set
$\{h(t,p)\}_{\forall p\in E}$ is total in $H_0$.
This means that if for some $F\in L^2(T,dm(t))$ the
following equation holds:
$$ 0=\int_Th(t,p)\, F(t)dm(t)\qquad \forall p\in E,
   \tag2.2 $$
then $F(t)=0$.

Let us assume that $L$ is injective. The operator $L^\ast:H\to H_0$
acts by the formula: 
$$(LF,g)_H=(F,L^\ast g)_{H_0},$$ thus
$$ L^\ast g=\int_E h(t,p) g(p)dp.
   \tag2.3 $$

Recall that we assume in this paper that $K$ and $L$ are injective,
so that $K^{-1}$ and $L^{-1}$ exist.
Let us state a simple lemma.

\proclaim{Lemma 2.1}
One has
$$ [LF,LG]=(F,G)_{H_0},
   \tag2.4 $$
provided that RKHS $H_K$ is defined by the kernel
$$ K(p,q):=\int_T\overline{h(t,p)} h(t,q)dm(t).
   \tag2.5 $$
\endproclaim

\demo{Proof}
One has
$$ [LF,LG]=(K^{-1}LF,LG)=(L^\ast K^{-1}LF,G)_{H_0}=(F,G)_{H_0},
   \tag2.6 $$
where the operator $L$ in (2.6),
after the first equality sign, is considered as an operator from $H_0$
into $H$. The last step in (2.6) is
based on the relation:
$$ L^\ast K^{-1}L=I.
   \tag2.7 $$
 Let us assume that $L^{-1}$ is a closed, possibly unbounded, densely
 defined operator from $R(L)\subset H$ into $H_0$, where $R(L)$ is the
 range of $L$. Then formula (2.7) is equivalent to the relation:
$$ K=LL^\ast.
   \tag2.8 $$
Indeed, in this case one has:
$$ K^{-1}=(LL^\ast)^{-1}=L^{\ast-1}L^{-1},
   \tag2.9 $$
so that (2.7) and (2.8) are equivalent.

Note that under our assumptions 
about $L^{-1}$ the operator $(L^\ast)^{-1}$ does exist
and $(L^\ast)^{-1}=(L^{-1})^\ast$.

Let us prove that (2.5) is equivalent to (2.8)
and, consequently, to (2.7). Using (2.1) and (2.3),
one gets:
$$ LL^\ast g=\int_T\overline{h(t,p)} \int_E h(t,q) g(q)dq\, dm(t)
   = \int_EK(p,q)g(p)dp,
   \tag2.10 $$
where $K(p,q)$ is defined by (2.5). Since $g(p)$ in (2.10) is
arbitrary, this formula implies (2.8), as claimed.
Therefore (2.5) implies (2.8), and, consequently, (2.7),
and (2.7) implies (2.4) according to (2.6).
Lemma 2.1 is proved.
\qed
\enddemo

In [Sa1] it is proposed to characterize the range $R(L)$ of
linear map (2.1) as the RKHS with the reproducing kernel (2.5).
It follows from Lemma 2.1 that if one puts the inner product
(1.1) of $H_K$, with $K(p,q)$ defined in (2.5), on the set $R(L)$,
then $L:H_0\to H_K$ is an isometry (see (2.6)).

In general one cannot describe the norm in $H_K$ in terms of some
standard norms, such as the Sobolev norm. 

{\it Therefore
the above observation (that $R(L)=H_K$ if one puts the norm of $H_K$
onto $R(L)$) does not solve the problem of characterization
of the range of $L:H_0\to H_0$ as an operator from $H_0$ into $H_0$.}

This point was discussed in [R2]. On the other hand,
some cases are known when one can characterize the norm in $H_K$ in
terms of the Sobolev norms (positive or negative) [R1].

It is also claimed in [Sa1,2] that an inversion formula exists for
a general linear transform (2.1) ([Sa2, p.56, formula (31)]).

This inversion formula is derived under 
the assumption [Sa2, p.58]
that $H_K$ is the space $L^2(E,d\mu)$, where $d\mu$ is some positive
measure. This assumption means that the 
kernel $A(p,q)$ of the operator
$K^{-1}$ is a distribution of the form $\delta(p-q)w(p)$, 
where $w(p)$ is the density of the measure $d\mu(p)$, 
that is $d\mu(p)=w(p)dp$, and 
 $\delta(p-q)$ is the delta function.

This and the definition of the inverse operator,
namely $KK^{-1}=I$, written in terms
of kernels, imply:
$$ \delta(p-q)=\int_E\delta(p-s) K(s,q) d\mu(s) =K(p,q) w(p),
   \tag2.11 $$
where we have assumed that  $w(p)>0$ is a smooth
function, with $v(p):=\frac{1}{w(p)}>0$. Thus (2.11) implies
that the reproducing kernel $K(p,q)$ must be of the form:
$$ K(p,q)=v(p)\delta(p-q),
   \tag2.12 $$
if one assumes that the inner product in $H_K$ is the same
as in $L^2(E,d\mu)$, as
indeed S.Saitoh assumes in [Sa2, p.56] and in [Sa1].

{\it Assumption (2.12) is not satisfied in general, and is essentially
equivalent to the formula $L^{-1}=L^\ast$, where $L$ now
is an operator from $H_0$ into $H_K$.}

Let us prove the above claim. If $L$
is considered as  operator from $H_0$ into $H_K$, then
 formula (2.7)  can be written as
$$ L^\ast L=I,\quad L:H_0\to H_K
   \tag2.13 $$
and formula (2.6) takes the form:
$$ ||LF||_{H_K}=||F||_0.
   \tag2.14 $$
{\it Thus $L:H_0\to H_K$ is an isometry (see (2.14)) and $L^\ast$ is the left
inverse of $L$ (see (2.13)).} 

We assume that $L$ is injective,
that is, the null-space of the operator $L$ is trivial:
$N(L)=\{0\}$. Since, by definition, $H_K$ consists 
of the elements of $R(L)$, that is, $R(L)=H_K$,
 and $L^\ast$ is injective on $R(L)$ by (2.13),
it follows that
$$ L^\ast=L^{-1},
   \tag2.15 $$
where $L^{-1}:H_K\to H_0$ is a bounded linear operator. 
The claim is proved. 

Formula (2.15) is equivalent to the inversion formula (31) in
[Sa2, p.56], while (2.14) is equivalent to formula (33) in [Sa2, p.57].

{\it It is now clear that the assumptions in [Sa1,2] are equivalent to the
assumption that $L:H_0\to H_K$ is a unitary operator, so that
its inverse is $L^\ast$.}

This assumption makes the description
of the range of $L$ and the inversion formula trivial.

It is suggested in [Sa1] and in [Sa2] to use the norm
$||f||_{H_K}=||L^{-1}f||_0 =||F||_0$ on $R(L)$,
where $L$ is an injective linear
operator, and it was claimed in these works
that one gets in such a way
a characterization of the range of the operator L defined 
by formula (2.1). In fact this suggestion
does not give a nontrivial and
practically useful characterization of the range $R(L)$ of this linear
integral operator because
the norm $||L^{-1}f||_0$ cannot, in general,
be described in terms of the usual norms,
such as Sobolev or Hoelder norms, for example.
Likewise, the fact that the inverse
of a unitary operator $L$ is $L^\ast$ does not give a
nontrivial inversion formula, since the main difficulty is to
characterize the space $H_K$ in terms of the usual norms
(such as Sobolev norms, for example) and to check that
$L:H_0\to H_K$ is a unitary operator.

Finally, one can easily check that if
the assumption in [Sa1, p.7] and [Sa2, p.56] holds (this assumption
says that $H_K$ has the inner product of $L^2(E, d\mu)$):
$$ \int_E\int_E A(p,q)f(p)\overline{g(q)} dp\,dq
   =\int_E f(p) \overline{g(p)} w(p)dp,
   \qquad \forall f,g\in H_K,
   \tag2.16 $$
where $A(p,q)$ is a nonnegative-definite kernel
of the operator $K^{-1}$ (see formula (1.1)), and $w(p)$ is
a continuous weight function, $0<c_0\leq\nu(p)\leq c_1$,
$p\in E$, then 
$$A(p,q)=w(p)\delta (p-q),
$$ 
which is an equation similar to (2.12), with $w(p)=v^{-1}(p)$.

This means that the assumption in [Sa2, p.56] that
the RKHS $H_K$ is realizable as $L^2(E,d\mu)$ is equivalent to
the assumption that the reproducing kernel $K(p,q)$ is of the form (2.12)
provided that $d\mu=w(p)dp$.

\vfill
\pagebreak

\vfill

\enddocument